\documentclass[preprint,showpacs,preprintnumbers,amsmath,amssymb]{revtex4}

\usepackage{amsmath}
\usepackage{amssymb}
\usepackage{graphicx}% Include figure files
\usepackage{dcolumn}% Align table columns on decimal point
\usepackage{bm}% bold math

\newcommand\A{{\mathcal A}}
\newcommand\C{{\mathcal C}}

\newcommand\Real{{\mathbb R}}

\newtheorem{thm}{Theorem}[section]
\newtheorem{rem}[thm]{Remark}

\newcommand{\de}{\partial}

\begin{document}

\preprint{APS/123-QED}

\title{A phenomenological approach to normal form modeling: \\
a case study in laser induced nematodynamics.} 

\author{ C. Toniolo, G. Russo, S. Residori and C. Tresser} 
           
\affiliation{C. To., G.R. and S.R.: INLN, 1361 Route des Lucioles, 06560, Valbonne, France.\\
C.Tr.: IBM, P.O. Box 218, Yorktown Heights, NY 10598, U.S.A.}

\date{\today}% It is always \today, today,

\begin{abstract}
An experimental setting for the polarimetric study of optically induced dynamical behavior in nematic liquid crystal films presented by G. Cipparrone , G. Russo, C. Versace, G. Strangi, and V. Carbone allowed to identify most notably some behavior which was recognized as gluing bifurcations leading to chaos. This analysis of the data used a comparison with a model  for the transition to chaos via gluing bifurcations in optically excited nematic liquid crystals previously proposed by G. Demeter and L. Kramer. The model of these last authors, relying on the original model for chaos by cascade of gluing bifurcations proposed by  A. Arneodo, P. Coullet, and C. Tresser about twenty years before, does not have the central symmetry which one would expect for minimal dimensional model for chaos in nematics in view of the time series near the gluing bifurcation. What we show here is that the simplest truncated normal forms for gluing with the appropriate symmetry and minimal dimension do exhibit time signals that are embarrassingly similar to the ones that could be found using the above mentioned experimental settings. It so happens that the gluing bifurcation scenario itself is only visible in limited parameter ranges, and that substantial aspect of the chaos that can be observed is due to other factors. First, out of the immediate neighborhood of the homoclinic curve, nonlinearity can produce expansion which easily produces chaos when combined with the recurrence induced by the homoclinic behavior. Also, pairs of symmetric homoclinic orbits create extreme sensitivity to noise, so that when the noiseless approach to attracting homoclinic pairs contains a rich behavior, minute noise can transforms the complex damping into sustained chaos. As Leonid Shil'nikov has taught us, combining global considerations and local spectral analysis near critical points is crucial to understand the phenomenology associated to homoclinic bifurcations in dissipative systems. We see here on an example how this helps construct a phenomenological approach to modeling experiments in nonlinear dissipative contexts. 
\end{abstract}

\maketitle

\bigskip
\section{Introduction}\label{sec:introduction}
\subsection{Foreword by C.Tr., for all the authors}
\emph{I was the first among us to meet Leonid  {\v S}il'nikov . It was in Tarusa, Russia, in  the summer of 1990 (one year before the generals' coup). We were participating to a very pleasant and friendly joint American-Russian conference which helped launch the journal CHAOS. Leonid's English was not enough to compensate for my 5 words rich Russian. Lerman and Afraimovich served as interpreters. I was quite impressed. This meeting  indeed only happened a few years after Arn{\'e}odo, Coullet, and myself first showed that the work of Leonid was instrumental in \textbf{understanding} many aspects of chaos (while most physicists were mislead to run around \textbf{measuring} dimensions and other often untrustworthy and prediction-void numbers). Leonid told me that when he saw our papers, he thought that the spirit of Poincar{\'e} was still alive in France, and added that given the out-of-mean-stream character of our approach, he thought of us as the 3 Musketeers.  I replied that he was d'Artagnan.  Laughters and  emotions were able to overcome the language barrier for a while, but to this day, I regret not having had a solid mathematical discussion with Leonid. His work has been important to Pierre Coullet and me both when we are working together and each in our own direction, to this very day, from normal forms to localized structures and open issues in reversible dynamical systems.  It is with great pleasure that, with my three younger collaborators and with Pierre Coullet who has helped us but decided that his contribution did not warrant co-authorship, we dedicate this paper to Leonid {\v S}il'nikov on the occasion of his $70^{\rm th}$ birthday. The influence of Leonid's work in this paper is quite clear. We hope that this contribution will be to his taste. }

\subsection{Back to business}
We reconsider here the problem of what should be the suitable models for certain experimental results on the dynamics of nonlinear interactions between a Nematic Liquid Crystal (NLC) and an intense optical field.  More precisely we are interested in models for the rich dynamics that was reported in \cite{Epx2Russoetal2000} using the experimental tools described in \cite{Exp1Cipparroneetal2000} in the case of an ordinary wave that impinges with a small angle on a homeotropically aligned nematic crystal film. A similar experimental setting was theoretically investigated in \cite{DemeterKramer}, that followed former experimental studies of such system (see for instance \cite{Exp-3Cipparroneetal1993, Exp-2Carboneetal1996, Exp-1Santamatoetal1998}  
and references therein). Then the results in \cite{Epx2Russoetal2000} were revisited in \cite{Epx3Carboneetal2001}  in view of  Demeter \& Kramer [1999]. \emph{Gluing bifurcations} (which were first systematically studied in \cite{CoGTGluingFirstnamed}, following the work in \cite{ACT} and \cite{GluingLorenzChaos1}) were recognized by Demeter and Kramer in the simplest non-trivial version of their Galerkin type truncation model, resulting from the coupling of the nematodynamics with the Maxwell equations \cite{DemeterKramer}. However, the symmetry of the model in  \cite{DemeterKramer}, as the symmetry in the first investigations of chaos by cascades of gluing  bifurcations in \cite{ACT}, and as the symmetry in \cite{KuramotoKoga} (a paper that appeared soon after \cite{ACT} where gluing was rediscovered independently), was an \emph{axial symmetry}. As we shall demonstrate on the basis of the experimental signals, this is not the proper symmetry to model the experiment (at least if one restricts oneself to the three main degrees of freedom).  The signal tells us that it is the \emph{central symmetry} that holds true in the experimental setting used in \cite{Exp1Cipparroneetal2000, Epx2Russoetal2000, Epx3Carboneetal2001}. We thus propose to model the experiments in \cite{Exp1Cipparroneetal2000, Epx2Russoetal2000, Epx3Carboneetal2001} using a new approach, the approach being in fact the main content of the paper. These experiences are collectively  called hereafter \textbf {the} \emph{laser induced nematodynamics experiment}: we understand that other experimental settings could be implemented, but only the one used in these three papers will be considered here.    

The modeling of the laser induced nematodynamics experiment will be done in a manner that can be considered as an extension to a richer case of the phenomenological protocol that is described below in one of the simplest cases (by which we mean ``one of the codimension one bifurcations").
This protocol is what we call \emph{phenomenological normal form based model building}. Rather than describing it abstractly, we have chosen to show it at work in a simple case. 

The simple case that we have chosen to present here to give an example of the protocol is characterized by \emph{the appearance of a limit cycle whose size grows from zero when a parameter is varied}. The protocol can then be described as follows:

1) {\bf Extracting gross features from the experiment:} We assume that we are confronted with a one parameter $R$ experiment where, as $R$ increases and some observable $x(t)$ is measured as times elapses for successive values of $R$, one gets the following sequence of asymptotic behavior for successive values of $R$ in the candidate phase plane with coordinates $(x(t), \dot{x}(t))$
(or, for instance,  $(x(t), {x}(t+\tau))$ for some suitably chosen time lapse $\tau$):
\begin{description}
\item[ $R<R_c$] The curve  $(x(t),\dot{x}(t))$ converges to a point $p_R$ that may either depend or not depend on $R$ (for instance, the point may be $(0,0)$ for all values of $R$ below $R_c$).
\item[ $R=R_c$] A transition value.
\item[ $R>R_c$] The curve  $(x(t), \dot{x}(t))$ converges to a \emph{closed curve} whose size grows with $R$, with the diameter going to zero as one approaches $R_c$ from above.
\end{description}

2) {\bf Proposing a normal form model in view of said gross features:} In view of 1), one proposes a model for the system at hand that reads:
\begin{equation}\label{eqn:Hopf}
\left\{ \begin{array}{lcl}    
			\dot{r} & = &\mu r  - r^3\\
                             \dot{\theta} &= &\omega_0 + \mu \omega_1
                             \end{array}
\right. ,
\end{equation}
where some readers will have recognized the lowest order truncation of the normal form for a so called \emph{``Hopf bifurcation"}, while in layman terms,  Eq. (\ref{eqn:Hopf}) is the simplest model for a simple one parameter bifurcation generating a stable cycle whose size increases from zero. Here $\mu$ stands for $R-R_c$, $r(t)$ represents $\sqrt{x^2(t)+y^2(t)}$ while $\theta$ is the angle defined by $x(t)+iy(t)=r(t)e^{i\theta t}$, and $\dot{\theta}= \omega_{0}$ at the bifurcation value $R=R_c$ of the parameter $R$.

3) {\bf Verifying predictions of the proposed model to validate it:} To check the validity of  Eq. (\ref{eqn:Hopf}), one measures the size of the radius of the limit cycle and the frequency as a function of $\mu= R-R_c$ and if:
 
\quad - the size grows as $\sqrt{\mu}$,  while 

\quad - the frequency of the cycle drifts linearly as $\mu$ increases above $0$,  

\noindent
one concludes that the model is satisfactory until a deeper, sound model based on the physics of the process under investigation is proposed, and one can meanwhile make predictions and try to use the model of Eq. (\ref{eqn:Hopf}) that can somehow be trusted to deduce it from the physics (or chemistry , or....) of the system at hand. 

There is one difficulty with this approach in that the domain of validity in $\mu$ of the model can be very small. This happens for instances in some relaxation systems that exhibit sigmoid shaped slow manifolds (for instance in variations of the Lenard equation). In such cases, the so called \emph{"canard"} effect (that has been studied in great details using  in particular  methods from non-standard analysis: see for instance \cite{BC, DD} and the related literature) makes limit cycle sizes jump abruptly by many orders of magnitude on tiny parameter intervals, sometimes very near the bifurcation point. However, one is probably quite safe in claiming that the above strategy works quite fine in the vast majority of systems that exhibit one parameter bifurcations to stable limit cycles... Anyway, it often works!  

How we propose to adapt this protocol to the more complicated context of the laser induced nematodynamics experiment will be the object of this paper:

- In Section 2, we will recall what was the main observable for the time signals extracted in the analysis of the experiment that we consider. We will also describe how to exploit the actual experimental time signals for three parameter values around the first qualitative change in the signal (that indeed corresponds to a gluing bifurcation) to collect the first information that we need. 

 - In Section 3, we will describe in words the model that we propose before we go to formulas. Next, we will write down the evolution equation corresponding  to the \emph{normal form  for $\zeta ^3$ with central symmetry} (in the language of \cite{AsymptoticChaos}:  $\zeta ^3$ means a generic codimension 3 bifurcation). As there are too many parameters in the full normal form, we will get a special truncated normal form for $\zeta ^3$ with central symmetry. Without trying to develop in a few lines the full theory, the normal form is obtained by keeping all resonant terms (that cannot be eliminated by change of variable) up to the first non-linear order, as determined by the set of eigenvalues with zero real part and the symmetry. In general, such full normal forms are not unique, as killing one resonant term by an appropriate change of variables may generate another one.  Anyway, few terms will be enough: more precisely we will provide the truncated normal form for $\zeta ^3$ with central symmetry that can also be obtained by \emph{scaling techniques}.  This is something that, making reference to \cite{AsymptoticChaos}  and \cite{CoulletSpiegel} (see also \cite{Elphick}), we like to call the \emph{asymptotic (normal) form (for $\zeta ^3$ with central symmetry)}.  Asymptotic normal forms happen to often capture what is most visible in the phenomenology (but in general not all of the phenomenology), that one can get when unfolding the singular situation when eigenvalues have zero real parts. In some cases, some important features need a fuller normal form, but a lot can already be said using just the asymptotic limit.
 
 - In Section 4, we review types of complicated behavior that can be related to isolated gluing bifurcations (as opposed to accumulation thereof).
 
 - In Section 5, we will show simulations for two model equations, and in parallel corresponding results of the laser induced nematodynamics experiment. The model with axial symmetry will be also used to help convince the reader of the better quality of the phenomenological model that we obtain here.

- Different issues, including:

\quad - what could go wrong in more traditional brute force approaches such as truncations of modes equations from complicated PDE's or even more complex evolution equations, and 

\quad -the nature of the chaos observed in the laser induced nematodynamics experiment,

\noindent
will be discussed in Sec. 6. 

Rather than describing the approach that we propose (both in term of dynamical signal treatment and model building) in the abstract (now that we have a simple codimension one example with a Hopf bifuration), we have chosen here to concentrate on an experiment that we consider rich enough to carry all that we want to tell. More than the choice of pedagogical point of view, it seems to us that the example that we treat (the laser induced nematodynamics experiment studied in \cite{Exp1Cipparroneetal2000}, \cite{Epx2Russoetal2000}, and  \cite{Epx3Carboneetal2001}) is enough of a challenge to allow us to claim the value of the approach on the basis of the success on this case.

\section{Extracting Gross Features of the Dynamics from the Time Signal}\label{sec:ExtractingTheDynamics}

\subsection{Time series for the laser induced nematodynamics experiment}\label{sub:TimeSeries}
 The time series observed in the work reported in \cite{Exp1Cipparroneetal2000}, \cite{Epx2Russoetal2000}, and  \cite{Epx3Carboneetal2001} for each value of the control parameter most often correspond to a quantity $\Theta(t)$, which is  \emph{the azimuthal angle of the major axis of the ellipse that represents the polarization state of the light trasmitted by the sample}.  
 The definition of $\Theta$ in terms of two of the Stokes parameters is mostly irrelevant for the rest of the discussion, but we have recalled it in Fig. 1. For more on the Stokes parameters and related matters, see \cite{RefPolar}.  See also \cite{Exp1Cipparroneetal2000, Epx2Russoetal2000, Epx3Carboneetal2001}  for the description of other signals extracted from the experiments. 
 
 \begin{figure}[htbp]
\centerline {\includegraphics[width=0.9\textwidth]{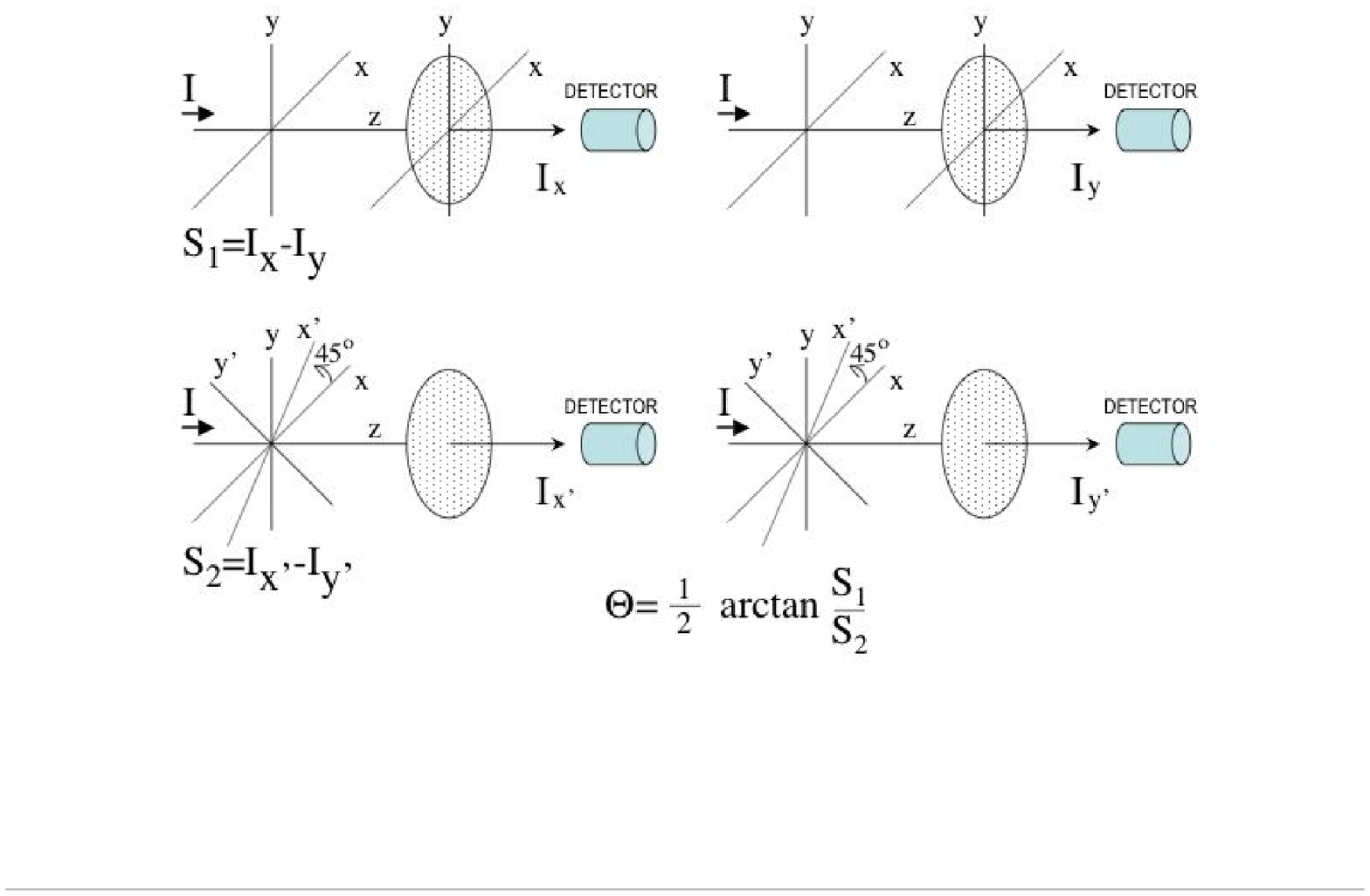}}
\caption{Definition of the two Stokes parameters $S_1$ and $S_2$, and of the azimuthal angle of the major axis $\Theta$.} 
\label{Fig1}
\end{figure} 

\subsection{Embedding dimension for the $\Theta$ time series}\label{sub:EmbeddingDimension}
We begin here our discussion of the treatment of the $\Theta (t)$ signal by discussing the phase space description extracted from time series. Among preferred methods for such extractions are:
 \begin{itemize}
\item Time delayed signals where for some $\tau$, one uses $(\Theta (t), \Theta (t+\tau), \Theta (t+2\tau), \dots )$.
\item The computation of successive derivatives, so that one considers  $(\Theta,\dot{\Theta},\ddot{\Theta}, \dots )$.
\end{itemize}

While the later is more noisy, given the difficulty of estimating a derivative from an already somewhat noisy primary signal, it offers the advantages of more likely match expectation with models obtained as normal forms.  This statement is due to the fact that successive derivatives appear in some truncated normal forms, and in particular in so called \emph{asymptotic normal forms}, on which we come back in the next section. Also, using time derivatives relieves one from having to choose an appropriate delay $\tau$.  The time derivative method is however not choice free, given the discrete nature of time sampling and the presence of noise: these two problems make one to choose an appropriate method to compute (rather to estimate)  the time derivative of the signals. In our case the derivative at a point is evaluated by interpolating to a second order polynomial, which uses the two forward and backward neighboring points, and then computing the actual derivative of the interpolating function. In order to get the best of two worlds, both methods of phase space description are shown for the phase space embedding of the the same three time series, obtained at three values  (1.6, 1.9, and 2.3) of the control parameter $\rho$ close to a gluing bifurcation that takes place at or close to $\rho=1.9$. We can see
the results of embedding the time series using respectively:

- the computation of two successive derivatives in Fig. 2 

- and delays by $\tau$ and $2\tau$ on the signal in Fig. 3.

\noindent
It turns out, as we will detail in the considerations to follow, that \emph{the choice is not relevant after all as far as extracting the main features of the signal is concerned}, a conclusion which one would  expect since the modeling is independent of the embedding method being used. However, when models will be proposed, it will turn out that the phase space evolution that they generate is closer to what one gets by using time derivatives. This preferred status of the time derivative method when it comes to details may be circumstantial and more precisely due to the fact that the best models that we get are so called \emph{control models} (models that come up in control theory), where the evolution equation relates several successive time derivatives of some quantity. 

The first element that we want to extract from the reconstruction is the \emph{dimension} of the phase space suitable for embedding the time signal obtained in the proper parameter range in the space $A^d$ (Fig. 2) with coordinates $(\Theta,\dot{\Theta},\ddot{\Theta},\dots )$ and in the space  $B_{\tau}^d$ (Fig. 3) with coordinates $(\Theta(t), \Theta(t+\tau), \Theta(t+2\tau),\dots )$ (both spaces are of course isomorphic to $\Real ^d$). 

In  Figures  2  and 3, where each panel shows the $d$-dimensional phase portrait for $d=3$ (in black) and the corresponding $d$ projections on the $(d-1)$-dimensional coordinate hyperplanes
projections (in red, green, and blue). The bottom projection, in red, corresponds to the embedding in dimension $d'=d-1=2$ (and so does the left vertical projection, in blue in Fig. 3 where one uses time delay with $\tau$, so that the right vertical projection, in green, corresponds to an embedding in dimension  $d'=d-1=2$ with  delay of $2\tau$). 

Comparing then the black and red displays in all panels indicates that clearly, dimension 2 is not enough for a phase space embedding while $d=3$ seems to be sufficient: thus we are inclined to accept $d=3$ as the appropriate embedding dimension throughout the parameter range that we are concerned with. A closer inspection is however appropriate before we make this important decision. 

Next, a simple inspection at Figs. 2-3 should convince the reader that, at the equilibrium point which is \emph{approached} by the observed orbits:

- there is an instability with a leading real eigenvalue, since the orbits that visit the equilibrium 
location get away from it  in an almost straight line (a feature which is mostly evident from panels (a), (b) of Fig. 3), 

-  the inward motion, being spiraling, is governed by a pair of complex conjugate eigenvalues which make 3 the minimal embedding dimension. 

\noindent
In fact:
  
- Some of the $(\Theta,\dot{\Theta})$  and  $(\Theta(t), \Theta(t+\tau))$ projections exhibit highly non-deterministic behavior in a two-dimensional phase space that would only be possible for a noise level in the signal much bigger that it really is. 

- To the contrary, the very cleanliness of the signal displayed in the spaces $A^3$ in Fig. 2,  (a) to (c), and even more $B_{\tau}^3$ in Fig. 3, (a) to (c), then indicates that:

\noindent
\textit{3 is indeed enough as an embedding dimension to capture the dynamics of the laser induced nematodynamics experiment.}

\begin{figure}[htbp]
\centerline {\includegraphics[width=\textwidth]{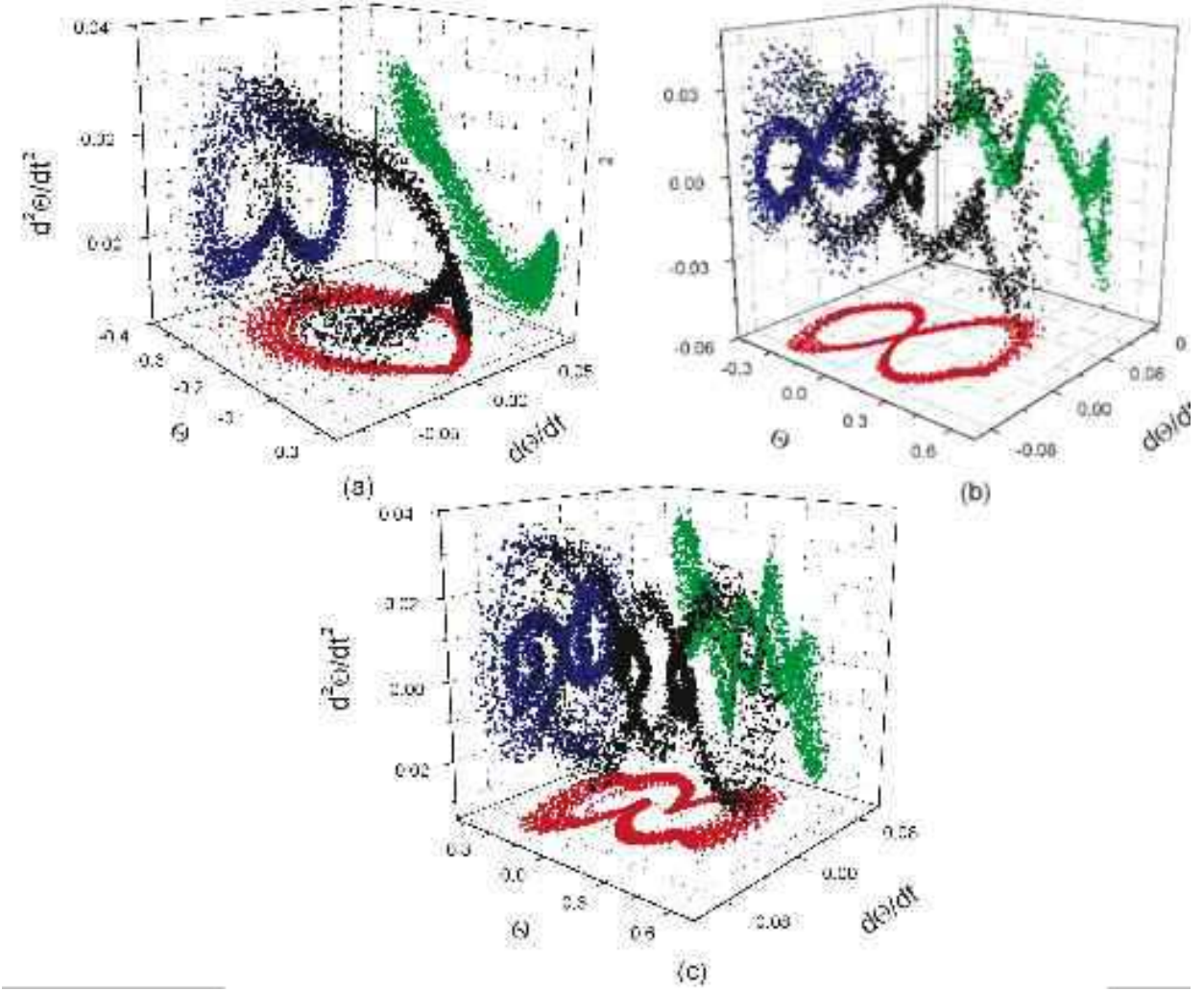}}
\caption{A gluing bifurcation in the $A^3$ space with coordinates $(\Theta, \dot\Theta, \ddot\Theta)$ of the signal $\Theta(t)$ produced 
             by the laser induced nematodynamics experiment. The time series in the three panels correspond to the ones 
             reported in  \cite{Exp1Cipparroneetal2000, Epx2Russoetal2000, Epx3Carboneetal2001} (and for this reason we
             refer to \cite{Epx3Carboneetal2001} for the definition of $\rho$), at three 
             different values of the control parameter $\rho$, specifically: (a) $\rho=1.6$; (b) $\rho=1.9$ and (c) $\rho=2.3$ 
             In (a) the parameter value is chosen before the cycle gets homoclinic; in (b) the parameter is just beyond the bifurcation 
             value for which gluing occurs, resulting in the creation of a bigger cycle, composed of two homoclinic trajectories; in
             (c) the parameter is taken farther from bifurcation, and the orbits leave the very proximity of the critical point $(0,0,0)$.} 
\label{Fig2}
\end{figure}

\begin{rem}
When we speak about approach to the critical point, we introduce a time element that is typically absent from figures in the embedding space, be it   $A^d$ or  $B_{\tau}^d$. However, ever if one has no access to the actual time evolution in a $d$-dimensional space, that allows to follow how the points are plotted one after the other with a good dynamic plotting routine, the information on what correspond respectively to the ``approach to" and to the ``escape from" the critical point can be guessed from (if not clearly read off) the time series.  
\end{rem}

\subsection{Observing a gluing bifurcation in the embedding space}\label{sub:GluingObservedInEmbeddingSpace}
In the embedding space, chosen either as $A^3$ or as $B_{\tau}^3$, one can observe a so-called \emph{gluing bifurcation} (as indeed predicted by Demeter \& Kramer [1999] on the basis of their model). Recall that a gluing bifurcation happens when two cycles have periods that diverge as the cycles approach a critical point when the parameter is close to the bifurcation value, under the following conditions and circumstances.

\noindent
\textbf{Conditions and circumstances for the gluing bifurcation:}

- \emph{ Two cycles approach a critical point when the parameter approaches the bifurcation value.}
 
 - \emph{ As the cycles approach the critical point the periods of these cycles diverge, as a consequence of the approach.}

- \emph{ The leading expansion at the critical point is weaker than the leading contraction, resulting in a bigger cycle being created after the bifurcation.}

- \emph{ The bigger cycle created by the bifurcation comprises and can be decomposed 
into approximate copies of the two pre-bifurcation cycles.}

\smallskip
\noindent
The gluing bifurcation is illustrated in Figs. 2-3, that show how a simple limit cycle ((a)-panel) gets glued to a symmetric one at the bifurcation value
((b)-panel), giving rise to a single bigger cycle with almost double period after the  bifurcation ((c)-panel). 
After cascades of gluing bifurcations leading to chaos in a system with axial symmetry were analyzed in \cite{ACT}, and similar bifurcations were discussed (slightly after but independently) in \cite{KuramotoKoga}, a systematic study of the gluing bifurcation was initiated in \cite{CoGTGluingFirstnamed}, where the role of different symmetries was indicated and the complexity away from the symmetric cases first suggested. In fact, as described in \cite{CoGTGluingFirstnamed},\emph{ gluing is a codimension two bifurcation among generic dynamical systems}, since in general one needs two parameters to bring two cycles to become homoclinic for the same parameter values. Clearly,  \emph{a symmetry turns gluing into a codimension 1 phenomenon}.  But a symmetry also kills the possibility of having cycles with complicated structure arbitrarily close to the bifurcation, in a neighborhood of the pair of homoclinic cycles (to the contrary of what could be called ``\emph{chaotic gluing}'', where the expanding eigenvalue is the biggest in absolute value) . The role of complex eigenvalues was examined in \cite{GGTGluingShilnik} and the theory further developed in \cite{GluingRenorm1} and \cite{GluingRenorm2} (see also \cite{Gambaudo} and references therein). 

\begin{figure}[htbp]
\centerline {\includegraphics[width=\textwidth]{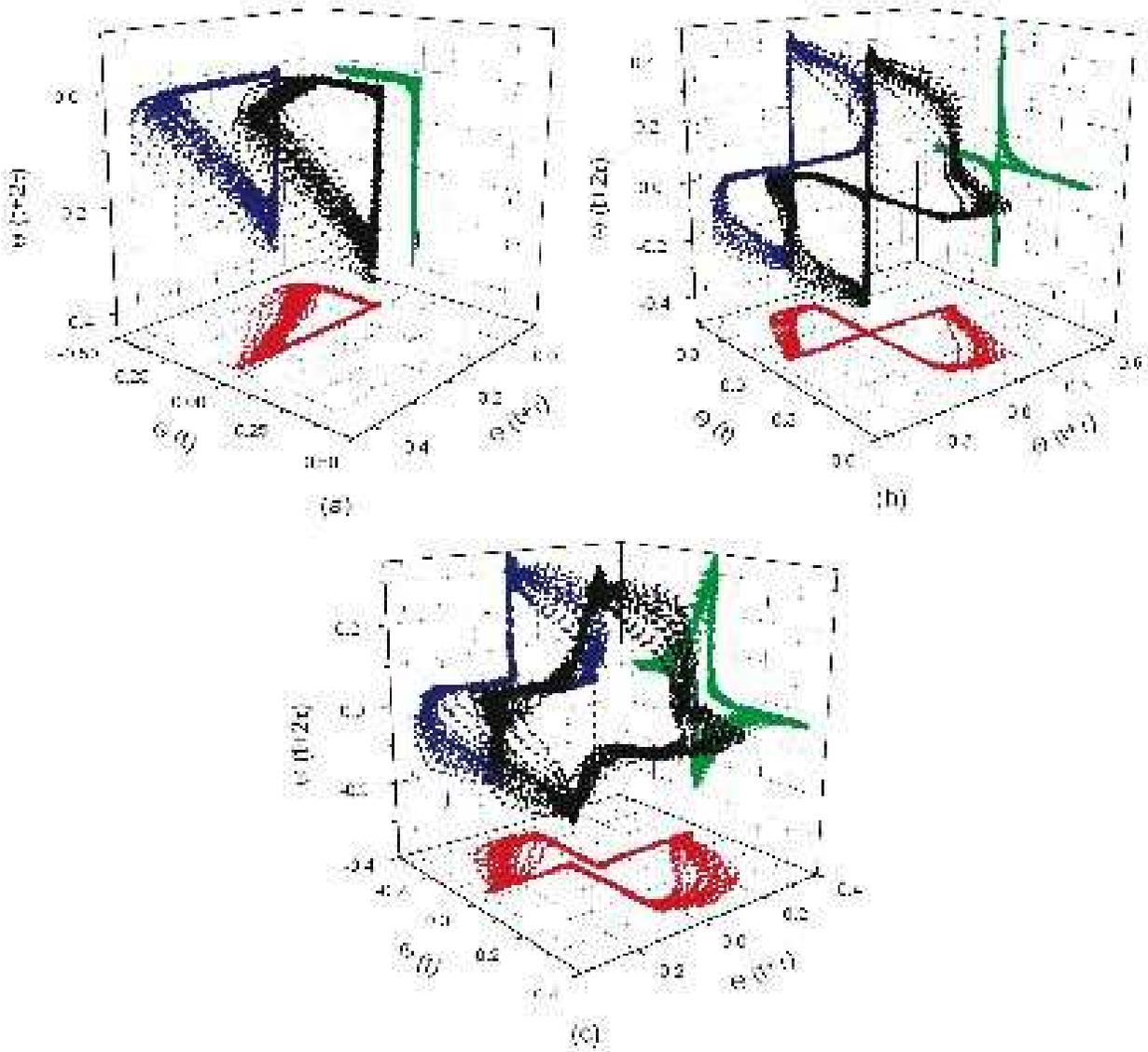}}
\caption{Representation in the $B_\tau^3$ space for the same experimental signal and parameter set as in Fig. 2. The panels from (a) to (c)
            show a gluing bifurcation of the time delayed signal in the space $(\Theta, \Theta(t+\tau), \Theta(t+2\tau))$, constructed by
            choosing a time lapse $\tau \sim 10 sec$ (corresponding with good approximation to the first minimum of the autocorrelation 
            function for the signal $\Theta(t)$).} 
\label{Fig3}
\end{figure} 

As already pointed out in \cite{Tresser}, the chaos in the case where all eigenvalues at the critical point are real is different from what one gets with a complex pair of stable eigenvalues: we will discuss that with more details below when we investigate the chaos of some $\Theta$ time series. We leave as an exercise to the reader to check that:  

\noindent
\emph{in three dimension, a pair of homoclinic orbits bi-asymptotic to the origin where the stable behavior is dictated by complex conjugate eigenvalues (the \emph{complex case}) is compatible with a central symmetry, but not with an axial symmetry}.  

\noindent
Recall that the axial symmetry is precisely the symmetry of the model discussed in \cite{DemeterKramer}. 

If one trusts the complex eigenvalues at 0, as suggested by the embedding of the $\Theta$ time series (to be more specific, as indicated by the bending 
of the reconstructed signal close to $0$ in Fig. 2 and in the panel (c) of Fig. 3), one can thus almost dispense of the symmetry seeking manipulation presented in the next subsection. We however present that investigation because in some cases, if not for the laser induced nematodynamics experiment, the nature or value of the eigenvalues may be hard to get (or at least to guess without tremendous experimental prowess).  For instance, one may have to deal with the problem of finding the symmetry (if any) in the \emph{real case}, by which we mean the case when all three eigenvalues at the central critical point are real.   
   
%Symmetries   
\subsection{Symmetries in the embedding space}\label{sub:SymmetryInEmbeddingSpace}
In the embedding space, again chosen either as $A^3$ or as $B_{\tau}^3$, one can observe, using parameter values after the gluing has occurred, 
as in Figs. 2-3 (b) and (c), that the system is invariant under the central symmetry $\C$ about the origin $0$, \textit{i.e.,} under the transform 
\[
\C: X \mapsto -X\,,
\]  
where $X= (x,y,z) \in \Real^3$ forms the coordinate system chosen for the embedding space. We notice that the central symmetry is particularly robust to coordinate change, which makes it easier to detect. 

\noindent
In fact, given the central symmetric system 
\begin{equation}
\dot X = L X + N(X)\,
\label{centralsymmetric}
\end{equation}
which we decompose into a linear ($L$) and a nonlinear ($N$) part, the system in Eq. (\ref{centralsymmetric}) is transformed under the change of variables $X = \Phi(Y)$ onto:
$$
\dot Y =  J^{-1} L\Phi(Y) + J^{-1} N(\Phi(Y)),  \qquad  {\rm where} \quad  J = \frac{\de \Phi}{\de Y}\,,
\label{centralsymmetric2}
$$
which has the same (central) symmetry of the original one if $\Phi$ contains only odd powers of $Y$ (\textit{i.e.,} $J$ is an even function of $Y$) . 

\emph{In particular, any linear change of variables respects the central symmetry.}

The same sort of conclusion fails to hold for a system invariant under an axial symmetry 
\[
\A:(x,y,z) \mapsto (-x, -y, z)\,
\] 
as for example in the original Lorenz model [1963] or the Lorenz-like model of \cite{ACT}, more relevant to our context:
\begin{equation}\label{eqn:Zeta3Symmetric}
\left\{ \begin{array}{l} \dot x = \alpha x - \alpha y  \\
                               \dot y = -4 \alpha y + x z + \mu x^3 \\
                               \dot z = - \delta \alpha z + x y + \beta z^2
          \end{array}
\right. ,
\end{equation}

Here, any linear change of coordinates that is not itself invariant under an axial symmetry ABOUT THE SAME AXIS destroys the original axial symmetry and deforms the structure of the attractor, partially hiding the symmetry. In some sense, the big difference between the case of $\C$ and the case of $\A$ is that in many cases, there is a single relevant center while all axes through that center play \emph{a priori} the same role. 

Notice that while the central symmetry of the time signals for the experiments that we consider is easy to recognize in $B_{\tau}^3$, it is obscured in the $(\dot{\Theta}, \ddot{\Theta})$ plane if one does not take the precaution of both getting as clean derivatives as possible, interpolating the time series. Blurring also happens if one use too long batches of time series at a time. 

\begin{figure}[htbp]
\centerline {\includegraphics[width=4in]{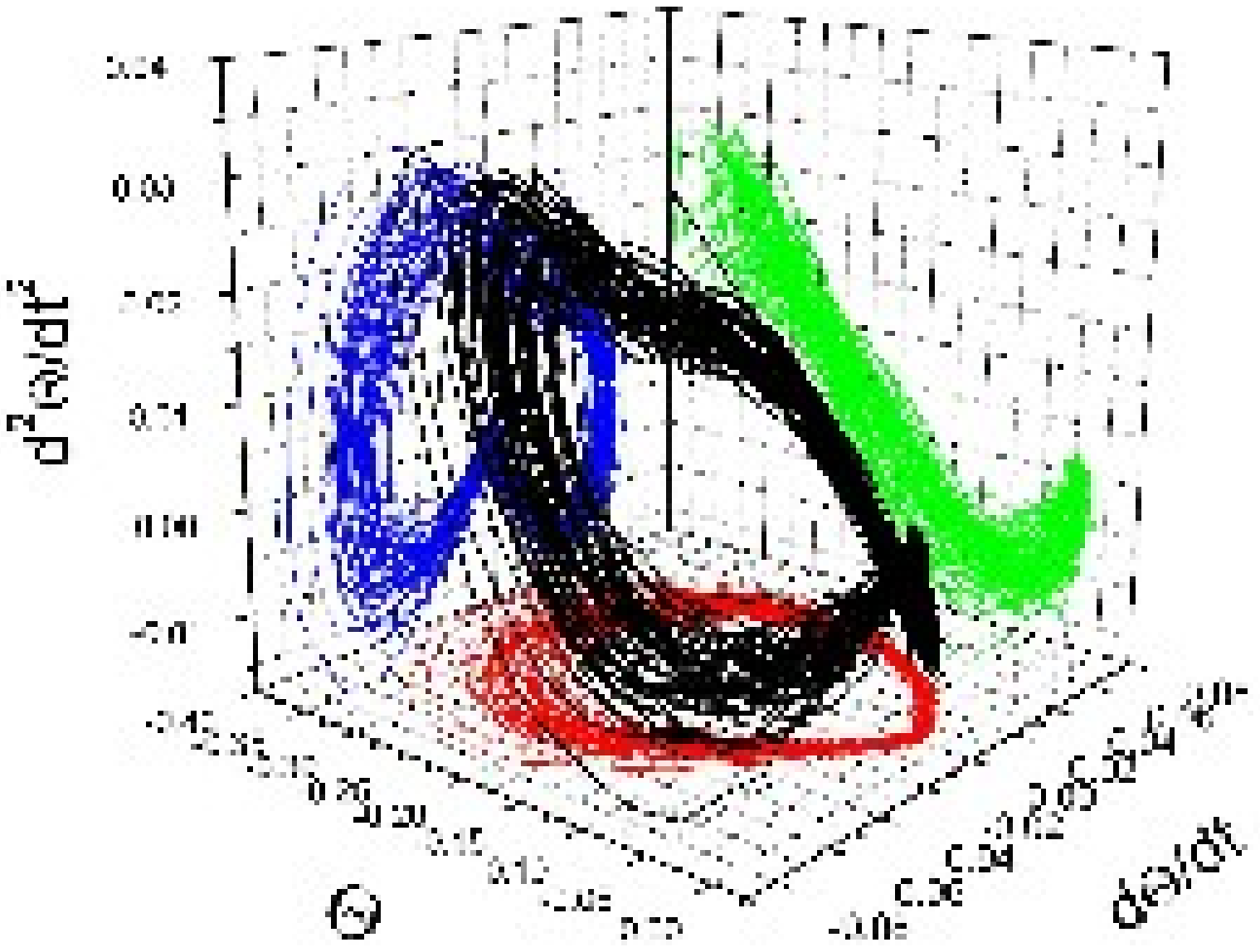}
                 \includegraphics[width=4in]{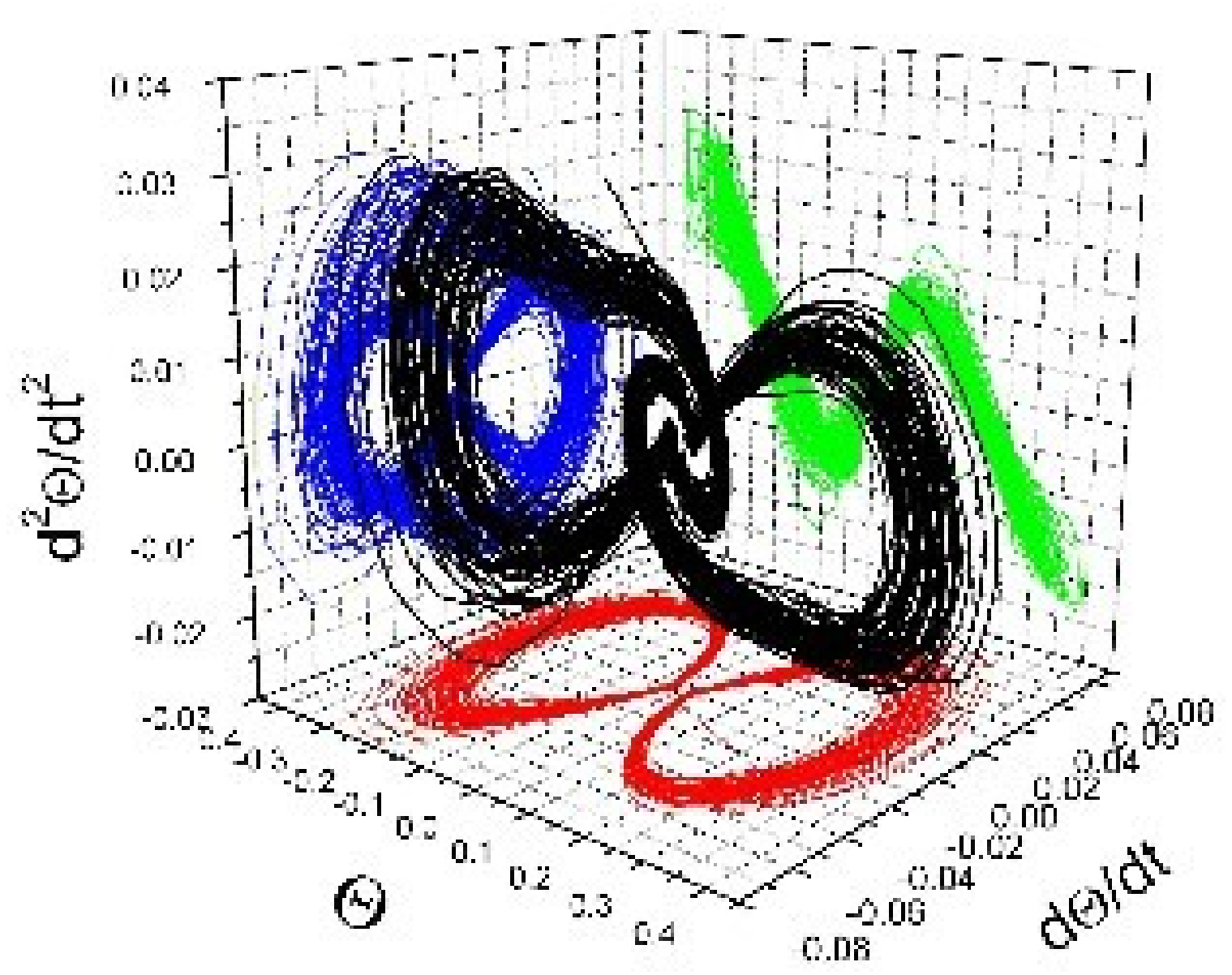}}
\caption{(a) Representation in the $A^3$ space for the same parameter set before the cycle gets homoclinic as in Fig. 2 (a), now using 
shorter time series for better visualization. In (b) the results of putting together the pre-homoclinic cycle in (a) with what is obtained in the same way from the time series $-\Theta(t)$.} 
\label{beforehomocl}
\end{figure} 

We point out that even before the gluing bifurcation, again shown in Fig. \ref{beforehomocl} (after which one can get in some sense from any of the two sides to the other), the central symmetry can in principle be detected. Ideally, this could be done as well as after the bifurcation if one would have a pair of  initial conditions that are images of each other under the symmetry (as one can do with numerical simulations). However, in many experimental settings, and the laser induced nematodynamics experiment is no exception, starting from initial conditions that lead to a pair of orbits that are symmetric of each other is hard (at best). 

So we want to indicate as a contribution to the toolkit that using both the recorded signal $\Theta(t)$ and the easily deduced signal $-\Theta(t)$ to perform embedding does produce two orbits that are exchanged by the actual symmetry, see Fig. \ref{beforehomocl} , assuming that the coordinates allow to see that symmetry at all (something which, as we have already indicated, is more easily realized for central symmetry than for an axial or planar symmetry, but for that matter the post critical parameter values do not make life easier): we invite the reader to compare Figs. 4-a and 4-b as a help to guess the spectrum of the linearized equation near the central critical point. Our own opinion is that the spiraling is significantly more visible on 4-b than on 4-a, a remark that may prove useful when studying other systems.

\begin{rem}
We use this opportunity to stress that too many authors, including distinguished colleagues, have overlooked the value of searching for symmetry in dynamical systems, and even overlooked the frequent essentiality of respecting the symmetry once they were recognized, \textit{e.g.,} when extracting Poincar{\'e} maps from the embedded phase portrait (despite the deep understanding that physicists often have of the importance of symmetry out of nonlinear physics).
\end{rem}

\section{Using the Gross Features of the Dynamics to Devise a Model}\label{sec:DevisingTheModel}

The observation of the gluing bifurcation in an embedding dimension 3, in particular with spiraling approach toward the critical point, and the observation of the central symmetry lead to two very simple assumptions:

- 1) For some set of three control parameters, say $\eta$, $\nu$, $\lambda$ there are simultaneously three coupled neutral modes, with generic linear coupling so that, at the triple singularity and in the space spanned by the modes, the matrix representing their interaction reads:
$$
L_{0}=\left(%
\begin{array}{ccc}
  0  &  1  & 0 \\
  0  &  0  &  1\\
  0  &   0  &  0 \\
\end{array}%
\right)\,,$$ 

- 2) The evolution equations are invariant under the central symmetry.

The idea is that the dynamics of an $n-$dimensional system,  with $n \geq 3$, close to the 
hyper-surface where three eigenvalues of the linear theory vanish, can 
be described even for complex and chaotic scenarios only through the marginal modes.
We can formalize the problem by considering a parameter dependent Ordinary Differential Equation of the form:
\begin{equation}
\frac{\de \tilde X}{\de t} = F(\tilde X, \mu)
\label{ODE}
\end{equation}
where $\tilde X \in \Real^3$, time $t \in \Real$, $\mu=(\mu_{1}, \mu_{2}, \cdots, \mu_{m})$ gives the set of parameters and 
$F$ is regular with respect to $(\tilde X, \mu)$. Given the linear part $L_{0}$ of the 
vector field, under certain assumptions \cite{Elphick} we look for a normal form of Eq. (\ref{ODE}) by performing a nonlinear change of variables:
$$
\tilde X = X + P_{\mu}(X) +O({\parallel X \parallel}^k)\,, 
$$
where $P_{\mu}(X)$ is a  polynomial of degree $k$  characterized by:
$$
DP_{\mu}(x) L_{0}^{\ast}= L_{0}^{\ast} P_{\mu}(x)\,,
$$
$DP_{\mu}(x)$ is the derivative of $P_{\mu}$ at $x$ and $L_{0}^{\ast}$ indicates the adjoint of the linear operator $L_{0}$.

In the case a symmetry $\C: X \mapsto -X$ exists, such as for example the central symmetry around the origin $0$, the vector field $F$ commutes with the linear map $\C$: 
$$
F(\C[X]) = \C[ F(X)]\,,
$$
and the same properties must hold for the normal form (\textit{i.e.,} for $\C$ the polynomials $P_{\mu}(X)$ must contain only odd powers of the components of $X$).
 
Using the standard procedure described above, we get for the case under consideration the following, general, normal form truncated to the third nonlinear order:
$$
\dot X = L X + N(X)\,,
$$
where $X=(x,y,z)$ and $L, N$ are given by:
$$
L=\left(%
\begin{array}{ccc}
  \mu_{1}  &  1  & 0 \\
  \mu_{2}   &  \mu_{1}   &  1\\
  \mu_{3}   &   \mu_{2}   &  \mu_{1}  \\
\end{array}%
\right)\,,
$$ 
and
$$
N=\left(%
\begin{array}{l}
  \mu_{4} x^3 + \mu_{5} x ( y^2-2xz)  \\
  \mu_{4} x^2 y + \mu_{5} y ( y^2-2xz) + \mu_{6} x^3  + \mu_{7} x ( y^2-2xz)\\
  \mu_{4} x^2 z + \mu_{5} z ( y^2-2xz) + \mu_{6} x^2 y+ \mu_{7} y ( y^2-2xz) +\mu_{8} x^3 + \mu_{9} x ( y^2-2xz) \\
\end{array}%
\right)\,.
$$
The same results may be rewritten in a very concise way as a third order  Ordinary Differential Equation:
\begin{equation}
\stackrel{...}{x}+ a_{\mu}(x,\dot x, \ddot x)\ddot x + b_{\mu}(x,\dot x, \ddot x) \dot x + c_{\mu}(x,\dot x, \ddot x) x= 0,   \nonumber
\end{equation}
where $x \in \Real$ and the nonlinear coefficients $a_{\mu},b_{\mu}$, and $c_{\mu}$ depend on the nine parameters ${\mu_{i}}$.
The former equation, which is \emph{the central symmetric $\zeta ^3$ codimension 3  normal form} according to the nomenclature of \cite{AsymptoticChaos}, is still redundant to illustrate the occurrence of chaos in a system with a triple instability (at least until more precise data justify using all this richness 
of parameters), and gluing and chaotic dynamics are observed \emph{easily enough for parameter values which are  close enough to the conditions for a triple bifurcation}, to let us hope that asymptotic methods will be enough to get  a good enough reduced equation.
Hence, by choosing suitable scalings for time derivation and $x$:
$$
\frac{\de}{\de t} \sim \epsilon, \qquad and \quad x \sim \epsilon^{3/2}
$$    
letting $\epsilon$ be a sufficiently small parameter, we eliminate higher order terms and recover the following asymptotic normal form, which is \emph{the central symmetric $\zeta ^3$  asymptotic normal form}, as P. Coullet likes to call such normal forms simplified ``to the bone" (hence quite often to the essential, even if aspects of the dynamics can sometimes be erased by simplifying from the complete normal form: see  \cite{AsymptoticChaos} and references therein) :
\begin{equation}
\left\{ \begin{array}{l} \dot{x}  =  y\\
                            \dot{y} = z\\
                            \dot{z} = -\mu x-\nu y -\beta z+ x^3   \nonumber 
                            \end{array}
\right. ,
\end{equation}
or
\begin{equation}\label{eqn:Zeta3SymOneLine}
\dddot{x}+ \beta \ddot{x}+ \nu \dot{x}+\mu {x}=  x^3 \, .
\end{equation}

Rescaling $x$ ($x$ to $\sqrt{\mu}x$) and fixing the coefficient of $\dot{x}$  to 1 allows one to rewrite Eq.(\ref{eqn:Zeta3SymOneLine}) as the system that was studied by one of the authors over twenty five years ago  \cite{ACT0, Tresser}.  Thus we get the following 
differential equation:
\begin{equation}
\dddot{x}+ \beta \ddot{x}+  \dot{x}= \mu x (1 -  x^2)\,, 
\label{eqn:themodel}
\end{equation}
which describes a sort of forced oscillator and where $\mu$ is here taken as a real positive parameter while $\beta$ stands for the dissipation (rate of contraction of the volume in phase space) of the associated flow. Using this system, we can observe a rich dynamics of gluing bifurcations  by studying one-parameter sub-families that correspond for instance to constant values of $\beta$ (these families are loosely speaking  \emph{generic} in that they cross the line in parameter space corresponding to homoclinic behavior).

Such a one parameter family was used already in \cite{ACT0}, but for weak dissipation $\beta$, which allowed to get what the authors of \cite{ACT0} soon after recognized as one instance of Sil'nikov chaos (see \cite[Tresser 1983a,b]{ShilnikovSymetric1}).
%\cite{ShilnikovSymetric1}, \cite{TresserShil1}, \cite{TresserShil2}). 
Here to the contrary, we choose to have a big dissipation $\beta$ (both types of dissipations were considered in \cite{Tresser} which offered a classification of homoclinic behavior; see \cite{TresserHomoclinicClassification2} for a more accessible short abstract that also filled a hole in the classification of Tresser [1981]) so that when we come close to the homoclinic situation, the real positive eigenvalue $\lambda$ has smaller absolute value than the real part of the contraction complex eigenvalues 
$-\rho\pm i \omega$, so that instead of satisfying what is sometimes called the condition for ``THE" Sil'nikov bifurcation $\lambda>-\rho\,,$ we have:
\begin{equation}
\lambda < -\rho\,.
\end{equation}

\begin{rem} When one speaks of  ``THE" Sil'nikov bifurcation, such denomination hides the fact L.P. Sil'nikov has studied many important bifurcations besides what is reported in Shilnikov [1965] and [1970]. In particular -among many examples related to various types of dynamics- he has studied stable homoclinic bifurcations as for instance in Shilnikov [1966] and [1968], that are particularly important in our context, as well as chaotic homoclinic and non-homoclinic bifurcations.
\end{rem}

\section{Background on Chaos Associated in Some Way to Gluing}\label{sub:GluingRelatedChaos}

\subsection{Transition to chaos by a cascade of gluing bifurcations}\label{sub:GluingCascadess} 

In the first occurrence \cite{ACT} of gluing,  a paper that has been cited in the literature on the laser induced nematodynamics experiment, the focus was indeed on chaos induced by a cascade of succession of elementary bifurcations of types A) and B), where:

- A) A type A bifurcation is a spontaneous symmetry breaking bifurcation, where stable symmetric cycles loose their stability and  generate a pair of asymmetric stable orbits that are exchanged by the symmetry.

- B) A type B bifurcation is a gluing of the pairs of cycles created by type A to generate a new, more complicated, symmetric stable cycle that for a later parameter value looses its stability by a type A bifurcation.

The symmetry that is involved in the discussion above (both in points A and B) is an axial symmetry, which can happen when the eigenvalues at 0 are real in a gluing bifurcation.  The global bifurcations are more complicated in some other cases such as when the stable eigenvalues are complex conjugate. The transition to chaos in the main scenario discussed in \cite{ACT}, is by an accumulation of gluing bifurcations (and an accumulation as well of the symmetry breaking bifurcation, since they are intertwined with the  gluing bifurcations: see points A and B above). 

In particular, because of the results in Shilnikov [1966], [1968] (see also \cite{TresserShil2} for less stringent smoothness conditions), one expects no chaos near the gluing of the simplest cycles that are each on one side of 0 (to the contrary of more complex cycles that get away from 0 almost along both branches of the unstable manifold), even out of a neighborhood of the butterfly formed by the homoclinic pair, at least in the case there does exist an axial symmetry (up to a change of variables). In fact, the ``simplest cycles" qualification can be dropped as the shape of the cycles is irrelevant.  We have: 

\noindent
\emph{One does not expect chaos at or near gluing bifurcation parameters in some neighborhood of the union of the two homoclinic loops that get glued.}

\subsection{From Sil'nikiv no-chaos theorem to Holmes' weak chaos}\label{sub:Sil'nik-Holmes} 
In fact (as we know from work of Leonid {\v S}il'nikov: see Shilnikov [1966], [1968]),  when there is a stable homoclinic orbit (or a pair thereof as in the case of a gluing bifurcation, a situation that was not explicitly studied in {\v S}il'nikov's original work, but the no-chaos result holds as well then):

\emph{{\bf {\v S}il'nikov's no chaos Theorem  \cite{ShilSta1}, \cite{ShilSta2}, \cite{TresserShil2} .} There is no chaos in a neighborhood of a homoclinic orbit cycle bi-asymptotic to a critical point whose leading instability eigenvalue(s) is weaker than the stability eigenvalue(s). There continues to be no chaos in that neighborhood for nearby parameter values.} 

In particular, there is no chaos at and near the gluing bifurcation parameter, except in a very weak sense described in \cite{Holmes} at the bifurcation value: the use of the very word ``chaos" in the case covered by P. Holmes was indeed disputed in \cite{TresserShil2} (notice that the mistake in \cite{Holmes}  of using the Grobman-Hartman continuous linearization Theorem instead of a smoother linearization is minor and corrected in  \cite{TresserShil2}, and is anyway of a technical nature). The nice effect discovered by Holmes is that in the case of gluing with complex stable eigenvalue and in the presence of a (then necessarily central, as we claimed above) symmetry:

\emph{{\bf Holmes pre-asymptotic complexity theorem \cite{Holmes, TresserShil2}:} For any infinite sequence $\underline{s}=s_1s_2,...$ of 0's and 1's one can find an orbit in $U_{\epsilon}$  that closely follows the two homoclinic cycles $H_0$ and $H_1$ according to the succession indicated by the sequence $\underline{s}$, all orbits being forward asymptotic to the  figure eight-homotopic  set $H_0\cup H_1$.}

This holds despite the fact, by {\v S}il'nikov's no chaos Theorem, in any neighborhood $U_{\epsilon}$ of the pair of homoclinic cycles and close to the bifurcation value, there are at most two periodic cycles (and a cycle there, if any,  is necessarily stable).

\subsection{From Holmes' weak chaos to weak noise induced chaos}\label{sub:Holmes} 
In particular, despite the nice structure described in \cite{Holmes} (see also \cite{TresserShil2}), there is no chaos  in the asymptotic behavior and in particular \emph{no positive topological entropy}, but only a simple asymptotic state (that is either $H_0$, $H_1$ or the critical point $0$ or the union $H_0\cup H_1=H_0\cup H_1\cup 0$) for any initial condition in $U_{\epsilon}$. 

However, in the presence of weak noise, that destabilizes the simple asymptotic objects, one has something undistinguishable from genuine chaos, the main fact being that since infinitely of the richness described by the above claim disappears off the bifurcation parameter, the noise is more and more quasi-chaos inducing when approaching the bifurcation value from any side, and in fact along any path in the parameter space. While the above claims fail to hold true in the real eigenvalues case (the presence of symmetry is in fact immaterial in the complex case), noise may still provide the type of 0-1 chaos for other pair of stable homoclinic behavior as there is a possibility to jump from one basin to the other when the orbit comes close to the critical point, and this already happens in dimension two. The noise enhancing character of homoclinic pairs is indeed a fascinating phenomenon, but whose precise study is beyond our scope now. One important feature of the noise induced chaos when there are several homoclinic orbits bi-asymptotic to the same critical point is that the noise needed to generate chaos is arbitrarily small at homoclinicity, with a threshold increasing in a rather obvious way when getting away from the homoclinic situation. Furthermore, a finite size noise give a  range of existence for such chaos which makes it relevant to explain what one observes in experiments. 

\begin{figure}[htbp]
\centerline {\includegraphics[width=0.9\textwidth]{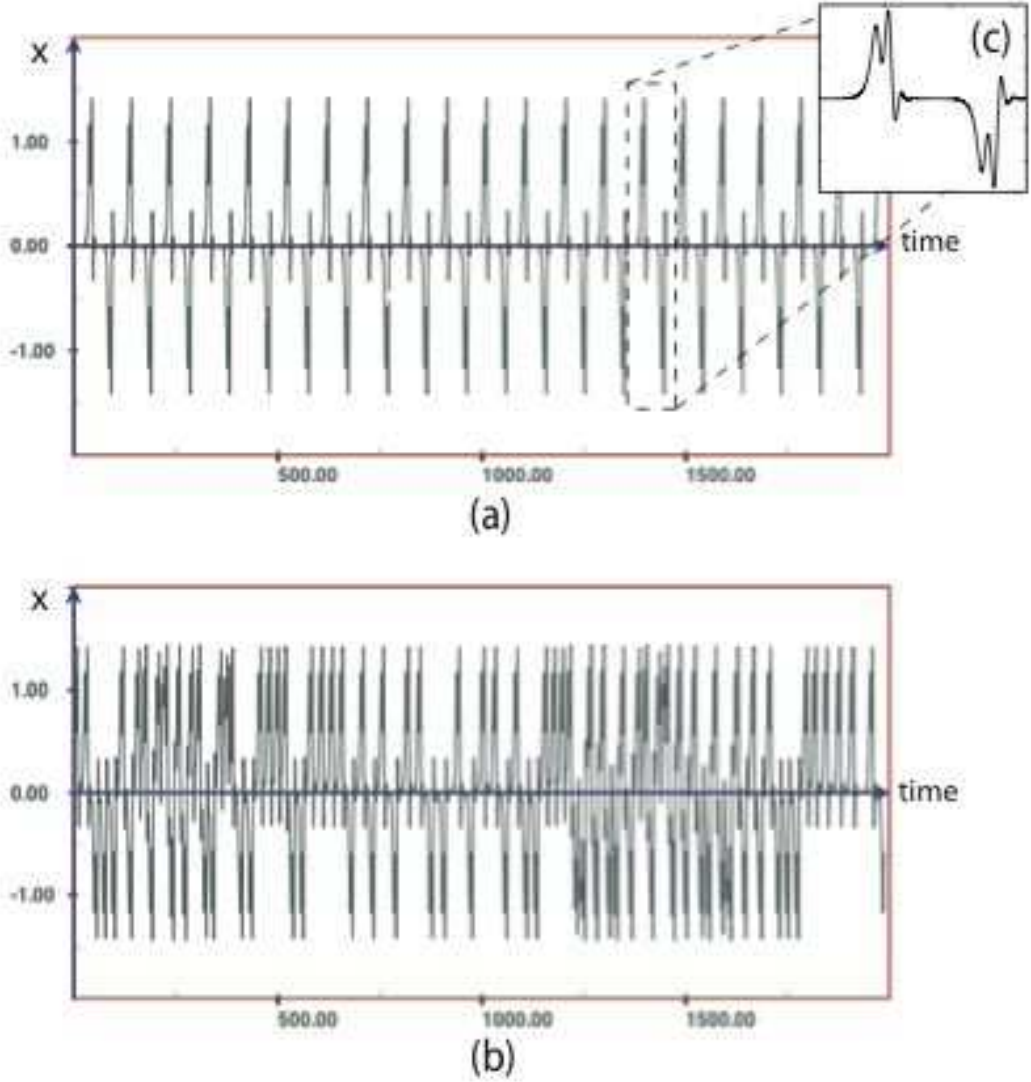}}
\caption{Holmes' weak chaos at the gluing of two homoclinic cycles, each making multiple loops around the critical point in $0$ as shown in
            the small inset (c). In the upper panel (a), a snapshot of the $x$-time signal from Eq. (\ref{eqn:Zeta3SymOneLine}), for values of the
            parameters $\beta=0.5$ and $\mu=0.631932480 \dots$ corresponding to gluing; the lower panel (b) reports a Holmes' like asymptotic 
            behavior for parameter values $\beta=0.5$ and $\mu=0.63195$, slightly before gluing.} 
\label{weakchaos}
\end{figure} 

By performing numerical simulations of Eq. (\ref{eqn:themodel}), we get what seems to be a behavior like the one described by Holmes at the gluing of two cycles, that each makes multiple loops in its basin, as illustrated in Fig. 5. Nonetheless, the huge number of digits that needs to be adjusted to get such a behaviour makes that a mathematical ``thing" rather than a physically relevant phenomenon (except of course if someone finds some application). 

\begin{rem} 
The Fig. 1-b, right column (model column) in \cite{Epx3Carboneetal2001} happens to be a preasymptotic behavior, as the comments in the text of that paper lets one guess. The asymptotic shape would be a regular alternation of ups and downs, but the aspect on Fig. 1-b left column (experimental time signal) in \cite{Epx3Carboneetal2001} would be recovered  by adding noise, as we will show in the figures presented in the next section.
\end{rem}
  
\subsection{Possible and probable chaos co-induced with homoclinic behavior}\label{sub:co-inducedChaos} 
To the contrary of such necessarily noise-induced chaos, as pointed out in \cite{Tresser}, one can expect chaos out of the neighborhood of the homoclinic orbits (with a signature on time signal quite different from the one obtained after a cascade of gluing bifurcations) in the case when there are complex conjugate stable eigenvalues. Rather than redeveloping the whole theory of first return maps in that case, whichever way the values of $\lambda$ and $\rho$ compare, we will start from the simplified approximate one dimensional return map, depending on a parameter $a$ that describes the departure from homoclinic behavior for both cycles (we here assume a symmetry as we have decided to remain close to the laser induced nematodynamics experiment). 

We notice that a single homoclinic orbit would lead to a map of the form
\begin{equation}\label{eqn:oneside}
x'= a+A  {\mid x \mid}^\delta \sin[\eta\, \ln(\mid x \mid + \phi)]\, +\,{\rm h.o.t.}, \qquad x\geq 0  \,,
\end{equation} 
so that in the symmetric case it becomes
\begin{equation}\label{eqn:co-inducing}
\left\{ \begin{array}{lcl} 
{x'} &=  a+ A_1  {\mid x \mid}^\delta \sin[\eta\, \ln(\mid x \mid + \phi_{1})]\, +\,{\rm h.o.t.}   \qquad x\geq 0\\
{x'} &= - a- A _2 {\mid x \mid}^\delta \sin[\eta\, \ln(\mid x \mid + \phi_{2})]\, +\,{\rm h.o.t.}  \qquad x<0
\end{array}
\right. ,
\end{equation}
where $\delta = \rho / \lambda < 1$ in our case, $\eta = \omega/\lambda$, $\phi_{1,2}$ are two arbitrary phases which, like the amplitudes $A_1$ and $A_2$, depend on the nonlinear part of the Poincar{\'e} map and the choice of the section where this map is considered. Recall that the one dimensional maps in Eqs. (\ref{eqn:oneside}) and (\ref{eqn:co-inducing}) are obtained from the actual (2-dimensional) Poincar{\'e} map by taking the limit of infinite dissipation (hence are only approximations, to the contrary of what happens in geometric Lorenz flows). The difference in the phases and in the amplitudes is absent if the ``surface of section" (where the first return map is the chosen Poincar{\'e} map) is chosen so that it respects the symmetry. In such case, the ``surface of section"  is made of two disjoint pieces, hence the quotation marks. Avoiding such disconnectedness obliges to abandon symmetry in the Poincar{\'e} map, hence in the one dimension limit of Eq. (\ref{eqn:co-inducing}).

When $\lambda <-\rho$ as is the case when considering gluing bifurcations, one thus sees ``why"
(but remember the approximate nature of the one dimensional return map approach, so one has not here a proof) no complicated asymptotic behavior occurs close enough to zero while the Holmes pre-asymptotic richness is rather transparent.  On the other hand, one can see that out of the range near 0 where $|x'|$ has to be smaller than $|x|$,  the derivative must have grown greater that 1 in absolute value so that one can have usual interval maps type complexity with period doubling type chaos, on any side of 0, and also an extra generator of complexity due to the possibility of getting to 
\[
x'<-x \qquad {\rm for} \qquad x>0\,,
\] 
and 
\[
x'>-x \qquad {\rm for} \qquad x<0\,,
\] 
that allows exchanging between the two sides orbits that stay bounded away from  zero. 

Coming back to the neigborhood of 0:

- When $a>0$, one has two stable periodic orbits for the flow close to homoclinicity, that correspond to the intersection on the graph with the diagonal. 

- The gluing is at $a=0$.

-When $a<0$, the graph cuts the second diagonal at symmetric points that correspond to the glued cycle.

\noindent
It is because such chaos comes from the same map of Eq. (\ref{eqn:co-inducing}) that describes the gluing that we say that \emph{the chaos is (possibly) co-induced with the gluing}. In the real eigenvalue case by contrast, the chaos is by cascades of gluing in the axial symmetric case as we have 
recalled from \cite{ACT}  (see also for instance \cite{GluingLorenzChaos1, GluingLorenzChaos2, GluingLorenzChaos3, GluingLorenzChaos4, GluingLorenzChaos5}) 

\section{Extracting Finer Features of the Dynamics from the Model and Comparing with 
the Experiment }\label{sec:FinerFeatures}
To check the validity of our approach, which basically starts from simple symmetry considerations 
and directly descend from them, we have  chosen to make a comparison between the signal from 
the laser induced nematodynamics experiment and the numerical simulation performed with:

-1) The central symmetric $\zeta^3$ codimension 3 reduced normal form given in Eq. (\ref{eqn:themodel}), which we will call \textit{Model $\C$} (as short for ``model with $\C$ symmetry"), and which we have obtained according to our protocol for normal form model building.

\noindent 
For the sake of comparison we also compare with the model proposed in Demeter \& Kramer [1999], more precisely:
     
-2) The Lorenz-like system with axial symmetry reported in Eq. (\ref{eqn:Zeta3Symmetric}), which we will call  \textit{Model $\A$} (as short for ``model with $\A$ symmetry").

\noindent 
More specifically, we will compare:

- The experimental nematodynamics signals of \cite{Epx3Carboneetal2001}, at four values of the control parameter $\rho_{E}$.

- Four simulations from Model $\C$ (given by Eq. (\ref{eqn:themodel})), three of them obtained by fixing $\beta_{C}$ and  varying   $\mu_{C}$ on a line, the last one at a point in the parameter set $(\beta_{C}, \mu_{C})$ not on the same line.  We want to point out that there is no reason for the parameters of our phenomenological model to coincide with the physical parameters  of the experiments; \emph{a priori}, we expect the parameter controlling the nematodynamics to rather be some nonlinear, complicated combination of $\beta_{C}$ and $\mu_{C}$, assuming (but this is a simplifying hypothesis whose validity or proper correction would need further study) that we have kept enough terms from the complete normal form.
  
- Numerical simulations from Eq. (\ref{eqn:Zeta3Symmetric}), where we fix respectively 
$\alpha_{A}=1.8$, $\beta_{A}=-0.07$, $\delta_{A}=1.5$ at four values of the control parameter $\mu_{A}$.  
 
We have added a small amount of noise to the numerical models, to obtain a more realistic behavior from an experimental point of view, and besides,  to observe the effect of small random fluctuations on our simplified models for the noiseless dynamics (this is of course much simpler than removing ALL the noise from the experiment). In fact, while some bifurcations such as Hopf bifurcations and period doubling bifurcations in natural system can be compared to what happens in noiseless experiments, homoclinic pairs act as noise amplifiers, which makes it more instructive to add an arbitrary small noise to the model. Of course, one may have to check that the phenomenology that one gets then is not dictated by the noise that has been chosen, or rather, one has to extract the features that are independent from the noise that has been chosen. 

\subsection{Extracting finer features from the time series}\label{sub:DynamicsOfTimeSignal}
Following the habits formed over the years in the study of nonlinear dynamics, we start from 
the inspection of the time series, that illustrate the evolution as time elapses of some measured 
quantity (such as $\Theta(t)$), and contain all the information about linear and nonlinear correlations of the experimental data (to the difference of Fourier analysis,  that has to do just with the linear correlations of the signal).

Time series surely provide us with interesting information about the dynamics of the system 
that produces the signal, but plotting the signal versus time it is just one of the possible visualization methods that are available to study nonlinear dynamics.   For example, some local and global properties are much more detectable through the reconstruction of the signal in the embedding space. This was exemplified:

- By the analysis made of the main properties of the leading eigenvalues of the spectrum of the linearized equations near the critical point that is the center of the symmetry toward the end of Sec. \ref{sub:EmbeddingDimension} for an example of local analysis,

- In Subsections \ref {sub:GluingObservedInEmbeddingSpace} and \ref {sub:SymmetryInEmbeddingSpace} for global considerations.

\begin{figure}[htbp]
\centerline {\includegraphics[width=0.9\textwidth]{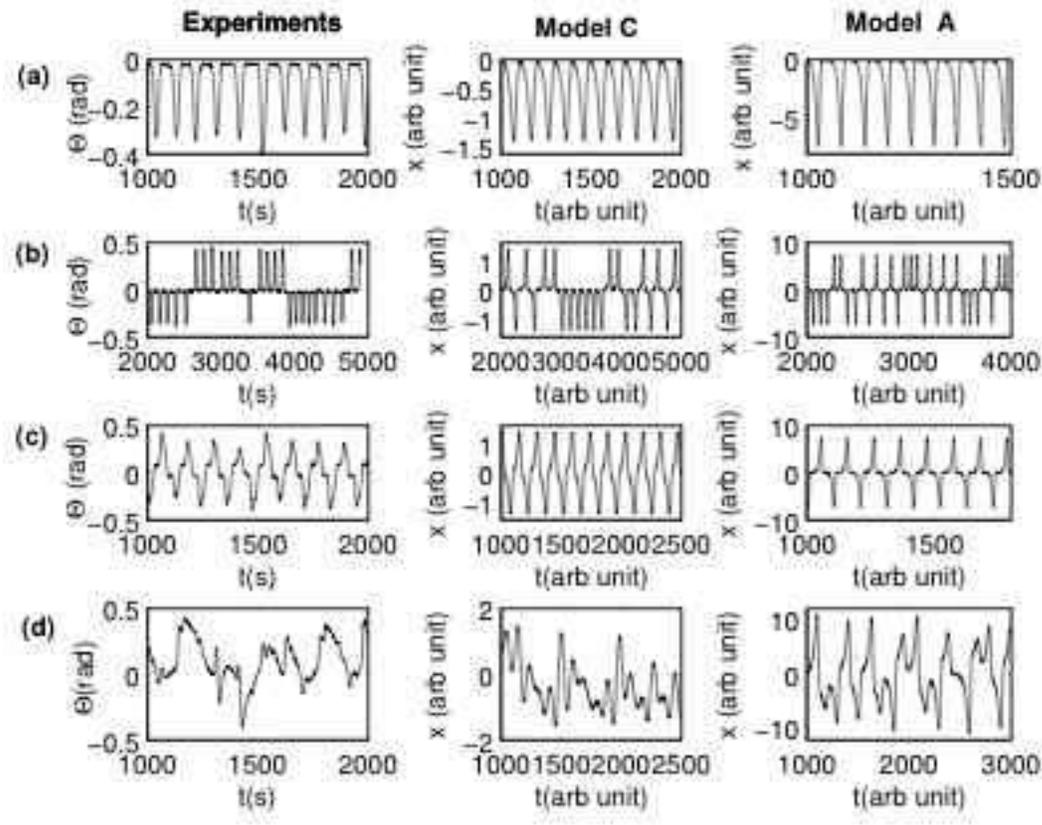}}
\caption{The time series: a gluing bifurcation and a ''disordered`` regime. In the figure, the left panels come
             from the experiments, the central column are the simulations from \textit{Model $\C$}, with central symmetry,
             while the right panels result from the \textit{Model $\A$}, with axial symmetry. In (a), an homoclinic limit cycle
             at the parameter values respectively of $\rho_{E}=1.6$; $\beta_{C}=2$ and $\mu_{C}=1.82$; 
             $\mu_{A}=0.0758$. (b) The weak noise induced chaos close to homoclinicity,  at  $\rho_{E}=1.9$; 
             $\beta_{C}=2$ and $\mu_{C}=1.8527$; $\mu_{A}=0.076$.  (c) A bigger cycle is created by gluing, with regular
             fluctuations which resist to noise  for  $\rho_{E}=2.3$;  $\beta_{C}=2$ and $\mu_{C}=1.865$; $\mu_{A}=0.0761$.
             In (d) an irregular behavior from the experimental signal and, besides, chaotic regimes from numerics, at parameter values
             of  $\rho_{E}=4.2$;  $\beta_{C}=0.2$ and $\mu_{C}=0.48$; $\mu_{A}=0.02$. The noise in the numerical simulations 
             (columns $\A$ and $\C$) is uniform in $[-0.02, 0.02]$.} 
\label{TimeSeries}
\end{figure} 

In Figures 6 to 9, what can be seen in the leftmost column of panels is from the experiment, what one sees in the central column is from the model that we have obtained in the present paper (Model $\C$), using symmetry and basic normal forms theory, and the rightmost column is from the model (Model $\A$) proposed in \cite{DemeterKramer}. Figure 6 represents the time series and Figs. 7, 8 and 9 respectively represent the projections from the three-dimensional phase space to the coordinate planes $XY$, $XZ$, 
and $YZ$. 

Let us now recall that only the gluing bifurcation, the fact that there are complex stable eigenvalues and a real expanding eigenvalue at the central critical point, and the symmetry were used.  This being said, we find the overall resemblance of what can be seen in the experimental (leftmost) columns to what one sees in the center columns of Figs. 6 to 9 generated from our model (Model $\C$)  almost embarrassing given the simplicity of the methods that we have used, but more of that in the next subsection.  Meanwhile, the accord between the simple patterns in the first and second column of 
Figure  \ref{TimeSeries} is nothing to tell the word about, since we mostly get back there what has been injected in the modeling process. In fact, the rightmost column, produced by Model $\A$, scores almost as well as the central column in Fig.  \ref{TimeSeries} from Model $\C$, except for a rather expert examination of  the three panels in the (c) line of Fig.  \ref{TimeSeries}. 

One line that appeal some special comments is line (b), where the chaos at or close to homoclinicity is due to the added noise in the center and rightmost columns: as discussed before, while the Holmes phenomenon may help the noise-sustained chaos, it is doubtful that pure deterministic dynamics be at the origin of the changes of basins observed in the experiment, given the difficulty to adjust parameters to get Fig. \ref {weakchaos}.  If no noise was used in the simulation of the models, the sequence of up (marked 1, say) and down (marked 0) would be a simple $01010101...$ instead of the complicated sequences that one gets by adding a small noise. 

As for line (d), while the chaos for Model $\A$ is what happens after accumulation of a cascade of gluing bifurcations that does not seem to take place for the laser induced nematodynamics experiment, what we get with Model $\C$ is of quite different nature. In fact, the parameters are so chosen that what we present in panel $\C$-(d) is an  example of what we called in Sec. \ref{sub:co-inducedChaos} ``chaos co-induced with the gluing'': we did not have to get this coincidence but wanted to keep the number of parameter values for Model $\C$ to 4. In fact, chaos co-induced with the gluing has so far not 
been reported in the laser induced nematodynamics experiment, but the phase space signatures corresponding to the experiment and to Model $\C$ are close enough.

\subsection{The dynamics in the embedding space}\label{sub:ComparWithExp}

\begin{figure}[htbp]
\centerline {\includegraphics[width=\textwidth]{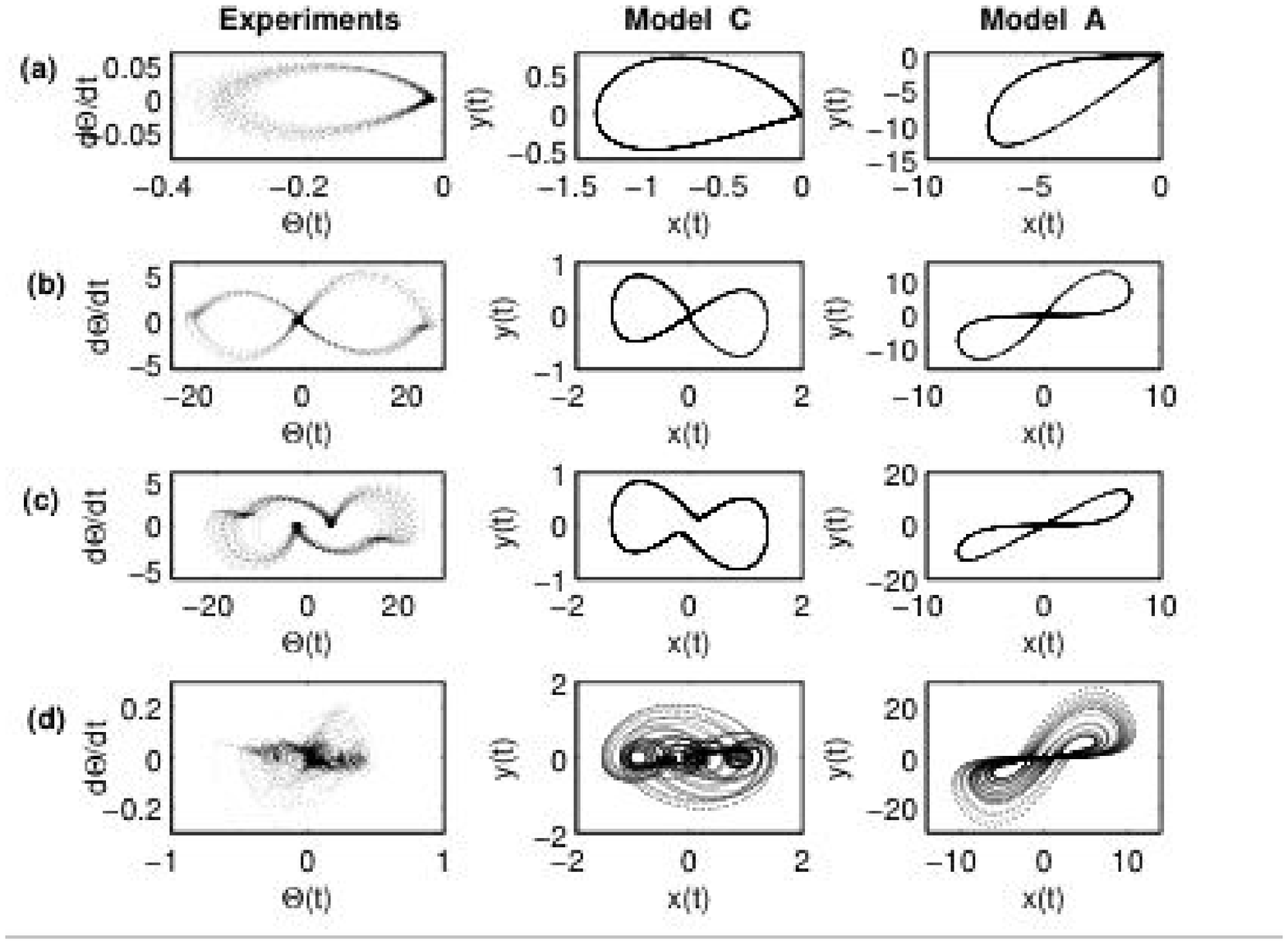}}
\caption{XY projections for the same parameters set of Fig. \ref{TimeSeries}.
            We find a clear agreement between the experimental measures and the predictions 
            of Model $\C$, while the agreement is not as good for Model $\A$ with a wrong symmetry. 
            The presence of complex eigenvalues in the experiments is evident if one looks at panels 
            (b), (c) and  also in (d), where from the relatively noisy signal we can extract the dynamics with 
            the noise filtered out sufficiently to allow  
            one to see the trajectory in phase space revolve around the axis
            $d \Theta/ dt =0$.} \label{FigXY}
\end{figure} 
 
\begin{figure}[htbp]
\centerline {\includegraphics[width=\textwidth]{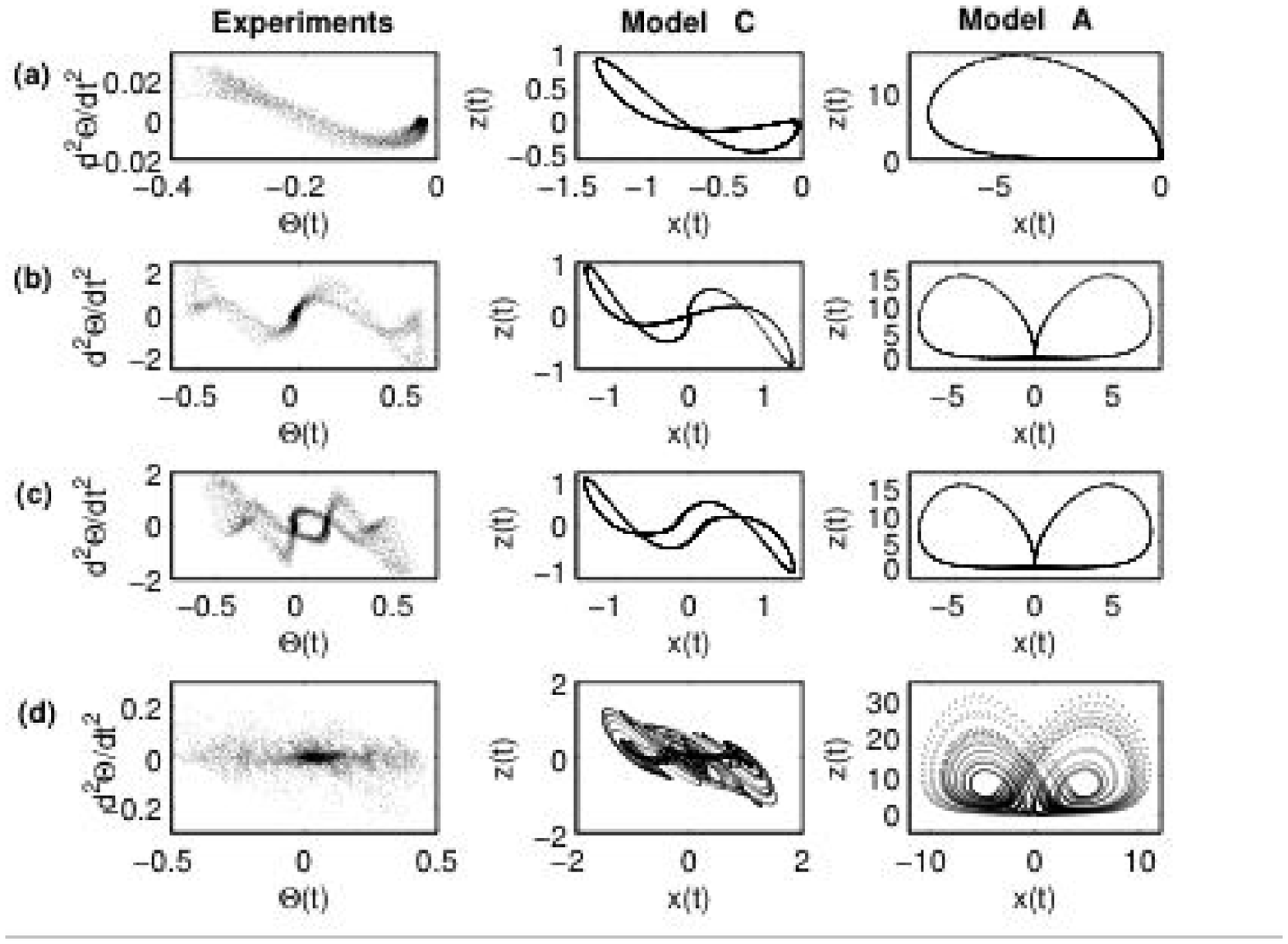}}
\caption{XZ projections  for the same parameters set of Fig. \ref{TimeSeries}.} \label{FigXZ}
\end{figure} 

\begin{figure}[htbp]
\centerline {\includegraphics[width=\textwidth]{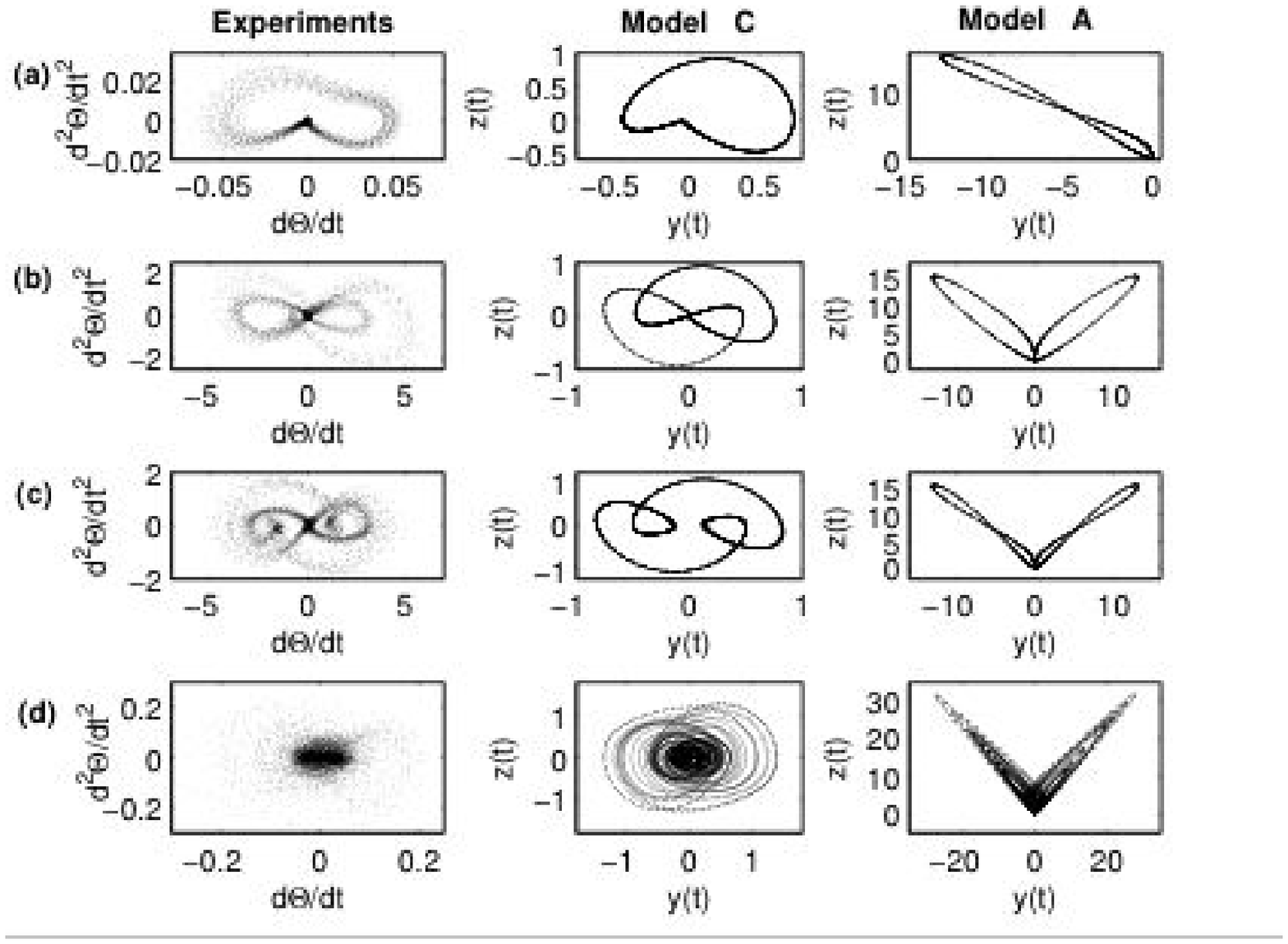}}
\caption{YZ projections for the same parameters set of Fig. \ref{TimeSeries}.} \label{FigYZ}
\end{figure} 

It is the comparison of the leftmost column to the central column in Figs. 7 to 9 that comes from simulations of Model $\C$ that motivates taking the pain of writing this paper,  and generates the hope that the protocol that we propose will be more broadly used.  The philosophy of this protocol can in some sense be understood as the collusion of the study of homoclinicity and symmetries in \cite{Tresser}  and \cite{TresserShil2} and of the use of normal forms in the way proposed in \cite{AsymptoticChaos}.  Pierre Coullet and too few others have used similar approaches, even if without explicitly proposing their success as coming from a very simple and quite general protocol as we have done here. We have to point that Coullet and his collaborators have also successfully used this protocol in the context of PDE for several years.

\section{Discussion and Some Remarks}\label{sec:DiscRem}

\subsection{On the axial symmetry in former models}\label{sub:FormerAxialSym}
Why would a method as standard as the one used in \cite{DemeterKramer} provide the wrong symmetry (especially in so expert hands)?  It has been known for long that normal forms on central manifolds should only be derived with the richness of the bifurcation near where they are justified. Galerkin type methods suffer the same constraint as any other method to get to reduced equations. A very well known example of an equation derived in a space with too many dimensions is provided by the Salzmann-Lorenz equations \cite{Sal}, \cite{Lor}. These equations are associated to the symmetry breaking bifurcation for the 0 equilibrium of the Boussinesque equations: 0 represents the homogeneous fluid and the critical points that get generated at the bifurcation are the cellular solutions, the symmetry between them being a reflection of the sense of the circulation in any chosen cell. This is a codimension one bifurcation, so that the central manifold has dimension one, which is of course enough for the symmetry breaking bifurcation normal form $ \dot{x}=\mu x-x^3$. This does not mean that a single mode should be kept, but that the modes one keeps should form the space in which the central manifold leaves, and only the normal form on that manifold makes sense. It was a very lucky circumstance that these equations, made famous by Yorke and then Ruelle, Williams, and Guckenheimer years after the masterful study made of them by E. Lorenz \cite{Lor}, contain a sort of non-hyperbolic chaos that is so rich and so mathematically instructive by itself. We notice that it was later realized that the Lorenz equations were themselves a truncated normal form, in fact in many important ways, but this would lead us way beyond our scope and we send the reader to the literature: see \cite{ShilShiTu}, \cite{PiShiTu}, \cite{CleCouTi} and several older references therein.  
  
\subsection{Value and limitations of the proposed model building protocol}\label{sub:ValueAndLimitOfProtoc}
Now that we have an equation that fits the actual behavior of the  laser induced nematodynamics experiment, we can hope that someone will deduce it from first principles to give meaning to the coefficients and compare with experiments at a quantitative level. One could meanwhile know whether the normal form truncation was well chosen by fitting the coefficients using experimental values at some special points such as the gluing bifurcation, to see if the concordance of behavior extends to events not taken into account in the fit. This might not enlighten us much on the physics until an actual theory is provided, but may help a lot... or can be misused .  

\section*{Acknowledgments.} 
We would like expecially to thank Pierre Coullet and Gerard Iooss for helpful discussions and remarks, which contributed to bring the paper to this final form. This work was partially supported by NSF Grant DMS-0073069, C.To. acknowledges financial support from the European Project PHYNECS (HPRN-CT-2002-00312), G.R. from the French Ministry of Research.
The simulations were done with the XDim Interactive Simulation package developed by Pierre Coullet and Marc Monticelli. 


\end{document}